\documentclass[12pt,preprint]{aastex}

\shortauthors{Mattson, Weaver \& Reynolds}
\shorttitle{\RXTE\ Seyfert 1 Observations}

\begin{document}

\newcommand{\RXTE}{\textit{RXTE}}
\newcommand{\ASCA}{\textit{ASCA}}
\newcommand{\suzaku}{\textit{Suzaku}}

\title{Possible X-ray diagnostic for jet/disk dominance in Type 1 AGN}

\author{Barbara J. Mattson\altaffilmark{\star}, Kimberly A.
Weaver}\affil{NASA/Goddard Space Flight Center, Astrophysics Science
Division, Greenbelt, MD, 20771}

\and
\author{Christopher S. Reynolds}\affil{Department of Astronomy,
University of Maryland, College Park, MD, 20742}

\altaffiltext{\altaffilmark{\star}}{also Department of Astronomy,
University of Maryland, College Park, MD and Adnet Systems, Inc.,
Rockville, MD}

\begin{abstract}
Using \textit{Rossi X-ray Timing Explorer} Seyfert 1 and 1.2 data
spanning 9 years, we study correlations between X-ray spectral features.
The sample consists of 350 time-resolved spectra from 12 Seyfert 1 and
1.2 galaxies.  Each spectrum is fitted to a model with an intrinsic
powerlaw X-ray spectrum produced close to the central black hole that is
reprocessed and absorbed by material around the black hole. To test the
robustness of our results, we performed Monte Carlo simulations of the
spectral sample.  We find a complex relationship between the iron line
equivalent width ($EW$) and the underlying power law index ($\Gamma$).  
The data reveal a correlation between $\Gamma$ and $EW$ which turns over
at $\Gamma \lesssim 2$, but finds a weak anti-correlation for steeper
photon indices.  We propose that this relationship is driven by dilution
of a disk spectrum (which includes the narrow iron line) by a beamed jet
component and, hence, could be used as a diagnostic of jet-dominance. 
In addition, our sample shows a strong correlation between $R$ and
$\Gamma$, but we find that it is likely the result of modeling
degeneracies.  We also see the X-ray Baldwin effect (an anti-correlation
between the 2-10 keV X-ray luminosity and $EW$) for the sample as a
whole,  but not for the individual galaxies and galaxy types.
\end{abstract}


\section{Introduction} \label{section:intro}

Time-resolved X-ray spectroscopy studies of active galactic nuclei (AGN)
offer the opportunity to investigate emission regions near the central
black hole.  In fact, X-ray spectroscopy offers the clearest view of
processes occurring very close to the black hole itself, probing matter
to its final plunge into the black hole.  Armed with such information,
we can unlock the structure of the innermost regions of AGN.

Typical X-ray spectra of AGN show an underlying powerlaw produced near
the central black hole with signatures of reprocessed photons often
present.  These reprocessed photons show up as an Fe K$\alpha$ line at
$\sim$6.4 keV and a ``reflection hump'' which starts to dominate near 10
keV and is produced by the combined effects of photoelectric absorption
and Compton downscattering in optically-thin cold matter irradiated by
the hard X-ray continuum.  The Fe K$\alpha$ line has been observed in
both type 1 (unabsorbed) and type 2 (absorbed) Seyfert galaxies.  It has
been attributed to either the broad line region, the accretion disk, the
molecular torus of unification models \citep{antonucci93}, or some
combination of these. Signatures of reflection have also been observed
in both Seyfert 1 and 2 galaxies.

If the unification models are correct, we should see similar spectral
correlations between Seyfert 1 and 2 galaxies, with any differences
easily attributable to our viewing angle. Regardless of the accuracy of
the reflection models, we expect changes in the underlying continuum to
drive changes in the reprocessing features.  However, results from X-ray
spectral studies of AGN have so far produced puzzling results.  Samples
of Seyfert 1 observations from \ASCA\ \citep{weaver01} and \textit{Rossi
X-ray Timing Explorer} \citep{markowitz03} have shown no obvious
relationship between changes in the continuum and iron line.  Several
galaxies have shown an anticorrelation between reflection and/or iron
line equivalent width and the source flux; e.g. NGC 5548
\citep{chiang00}, MCG $-$6-30-15 \citep{papadakis02}, NGC 4051
\citep{papadakis02, wang99}, NGC 5506 \citep{papadakis02, lamer00}. 
Recent data from \suzaku\ on MCG $-$6-30-15, on the other hand, show
that the iron line and reflection remain relatively constant while the
powerlaw is highly variable \citep{miniutti06}. \citet{zdziarski99}
found that Seyfert galaxies and X-ray binaries show a correlation
between the continuum slope and reflection fraction, so those with soft
intrinsic spectra show stronger reflection than those with hard spectra.
 However, other studies have found either a shallower relationship than
Zdziarski et al. \citep{perola02} or an anticorrelation
\citep{papadakis02, lamer00}.

Here we present the first results of a larger study of the X-ray
spectral properties of Seyfert galaxies observed by the \textit{Rossi
X-ray Timing Explorer} (\RXTE).  Our full study consists of observations
of 30 galaxies.  In this letter, we focus on the spectral results from
the subset of 12 Seyfert 1 and 1.2 galaxies.  In
\S~\ref{section:analysis} we present our method of data analysis,
including our sample selection criteria (\S~\ref{subsection:sample}), a
description of our data pipeline (\S~\ref{subsection:pipeline}), and
results of our spectral analysis (\S~\ref{subsection:results}).  We
discuss the implications of our results in \S~\ref{section:discussion}
and detail our conclusions in \S~\ref{section:conclusions}.

\section{Data Analysis} \label{section:analysis}

\subsection{The Sample} \label{subsection:sample}

The \RXTE\ public archive\footnote{Hosted by the High Energy
Astrophysics Science Archive and Research Center (HEASARC;
\url{http://heasarc.gsfc.nasa.gov/})} represents one of the largest
collections of X-ray data for AGN, with pointed observations of over 100
AGN spanning 10 years.  The \RXTE\ bandpass allows the study of
absorption and iron line properties of AGN spectra, as well as a glimpse
at the Compton reflection hump.  We use data from the \RXTE\
proportional counter array (PCA), which is sensitive to energies from 2
to 60 keV and consists of five Proportional Counter Units (PCUs).  Most
of the sources in our sample do not show significant counts in the
\RXTE\ Hard Energy X-ray Timing Experiment (HEXTE), so we do not include
HEXTE data in this study.

To focus this study, we choose only Seyfert galaxies for which the
\RXTE\ public archive contained a minimum of two pointings separated by
at least two weeks.  We further required the total observed time be
$>40$ ks.  These selection criteria led to a sample of 40 Seyfert
galaxies. For the analysis presented here, we examine the 18 Seyfert 1
and 1.2 galaxies. Six galaxies were eliminated after they were put
through our data pipeline (see \S~\ref{subsection:pipeline} for more),
so the final sample presented here consists of 12 galaxies, listed in
Table~\ref{table:sources}. Because the data come from the public
archive, the sample is not uniform from galaxy to galaxy or even from
observation to observation; however, we use the Standard 2 data, which
provides a standard data mode for these diverse observations.

\subsection{Data Pipeline} \label{subsection:pipeline}
 
To ensure consistent data reduction of the large volume of data, we
developed a data pipeline. The Standard 2 data for each observation was
reduced using a combination of FTOOLs and the Python$^{\circledR}$ scripting
language.  The pipeline produces time-resolved spectra, each with a
minimum of 125,000 net photons, which are extracted using standard PCA
selection criteria and background models \citep{jahoda06}.  Sources
which did not have sufficient net photons for even one spectrum were
eliminated from the final sample (Table~\ref{table:sources} shows the
final sample with the 6 eliminated sources listed in the table
notes).  Each spectrum includes 1\% systematic errors.  We are confident
in the instrument response and background models up to energies of
$\sim$25 keV, so we ignore channels with higher energies.

\subsection{Spectral Fitting and Results} \label{subsection:results}

The data pipeline produced 350 spectra for the 12 galaxies in our
sample.  Each spectrum was fitted from 3 to 25 keV with an absorbed
Compton reflection model plus a Gaussian iron line.  In \texttt{xspec},
the PEXRAV \citep{magdziarz95} model simulates the effects of an
exponentially cut-off powerlaw reflected by neutral matter and has seven
model parameters: photon index of the intrinsic underlying power-law
($\Gamma$), the cutoff energy of the power law in keV ($E_c$), the
relative amount of reflection ($R$), the redshift ($z$), the abundance
of heavy elements in solar units ($Z$), the disk inclination angle
($i$), and the photon flux of the power law at 1 keV in the observer's
frame ($A$). The relative amount of reflection is normalized to 1 for
the case of an isotropic source above a disk of neutral material
($\Omega=2\pi$). Adding a Gaussian line (energy in keV ($E_{Fe}$),
physical width ($\sigma$) in keV, and normalization in units of photons
cm$^{-2}$ s$^{-1}$) and an absorbing column ($N_H$, in cm$^{-2}$) yields
a total of 12 parameters.

We fixed the following values in PEXRAV: $E_c$ = 500 keV, $Z$ = 1.0, and
$\cos i$ = 0.95.  This inclination represents an almost face-on disk;
however, since we are seeking trends in the spectral parameters, rather
than absolute values, the precise value is not important to this study.
In addition, $z$ is fixed at the appropriate value from the NASA
Extragalactic Database for each
galaxy\footnote{\url{http://nedwww.ipac.caltech.edu/}}. After fitting
all spectra to this model, we derived the mean Gaussian width for each
source (Table~\ref{table:sources}), then held $\sigma$ fixed for a
second fit to the model.  Our final model has free parameters: $\Gamma$,
$R$, $A$, $E_{Fe}$, iron line normalization and $N_H$.  To prevent
\texttt{xspec} from pursuing unphysical values of the parameters, we set
the following hard limits: $0 \leq \Gamma \leq 5$, $0 \leq R \leq 5$, $
5.5 \leq E_{Fe} \leq 7.5$ keV, and $0 \leq \sigma \leq 1.5$ keV (for the
free-$\sigma$ fits).

\begin{figure}
\plottwo{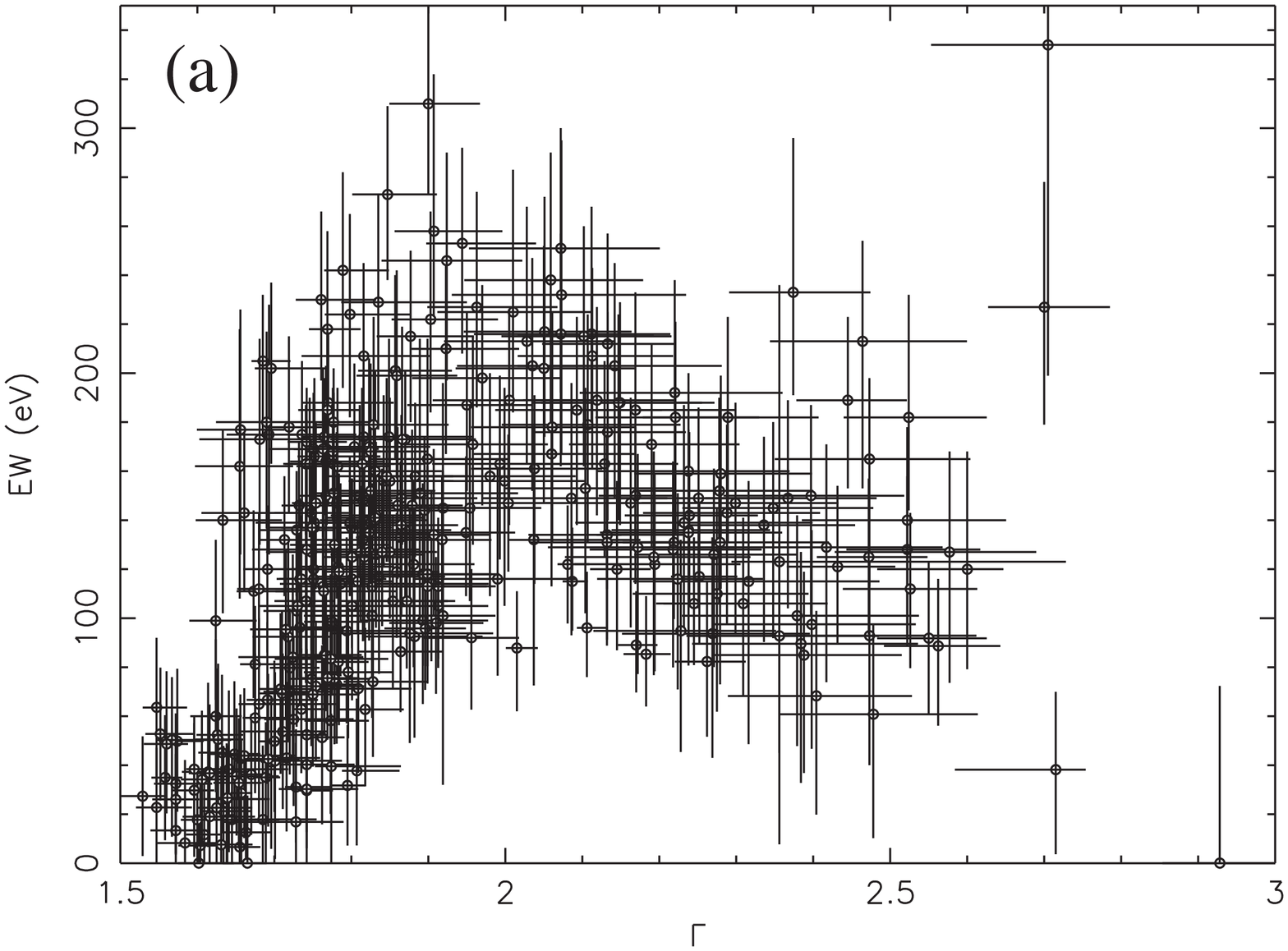}{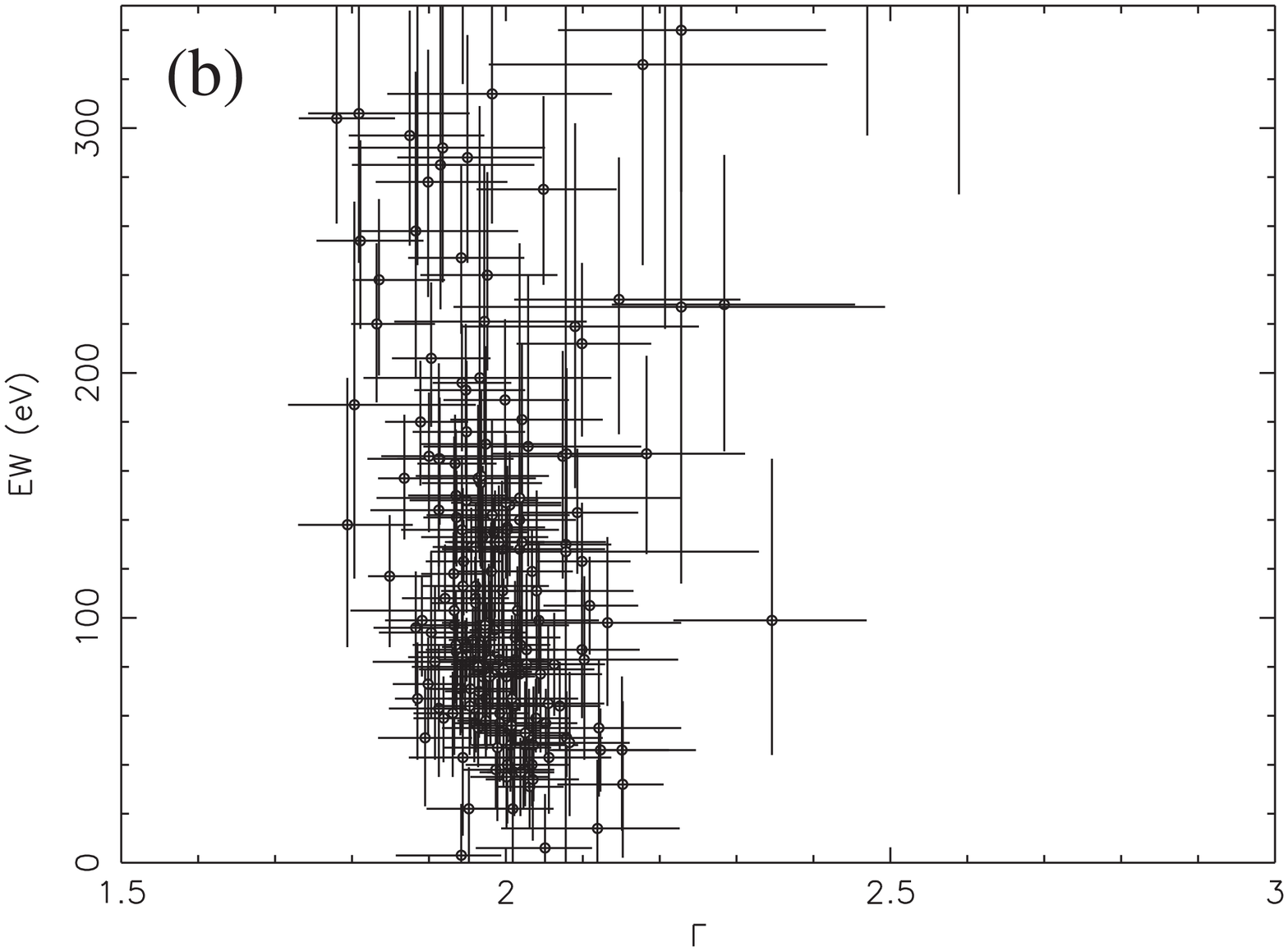}
\caption{Iron line equivalent width in eV ($EW$) versus powerlaw photon
	index ($\Gamma$) for the Seyfert 1/1.2 sample (a) and for the Monte Carlo
	simulations (b).}
  	\label{fig1}
\end{figure}

\begin{figure}
\plottwo{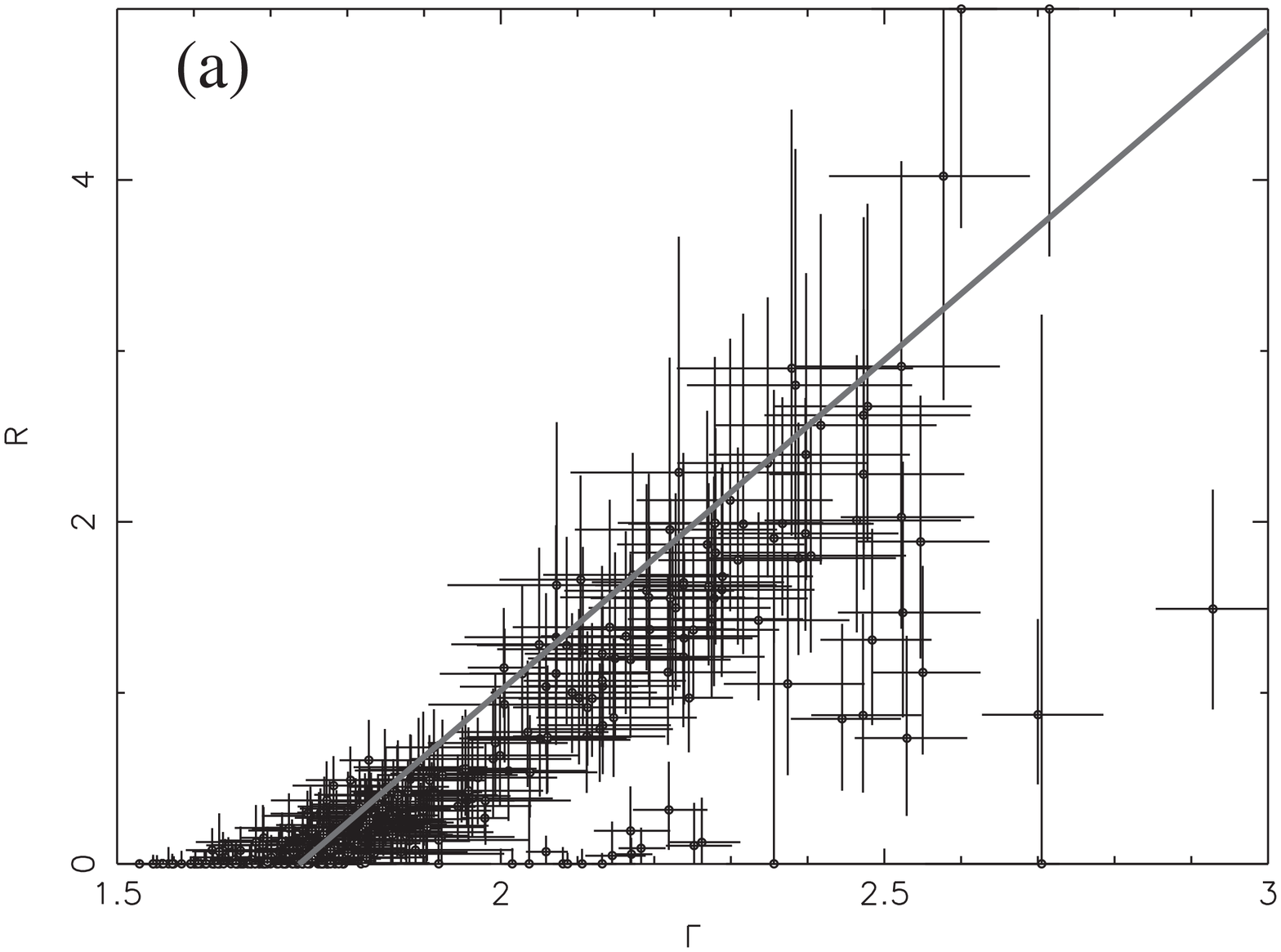}{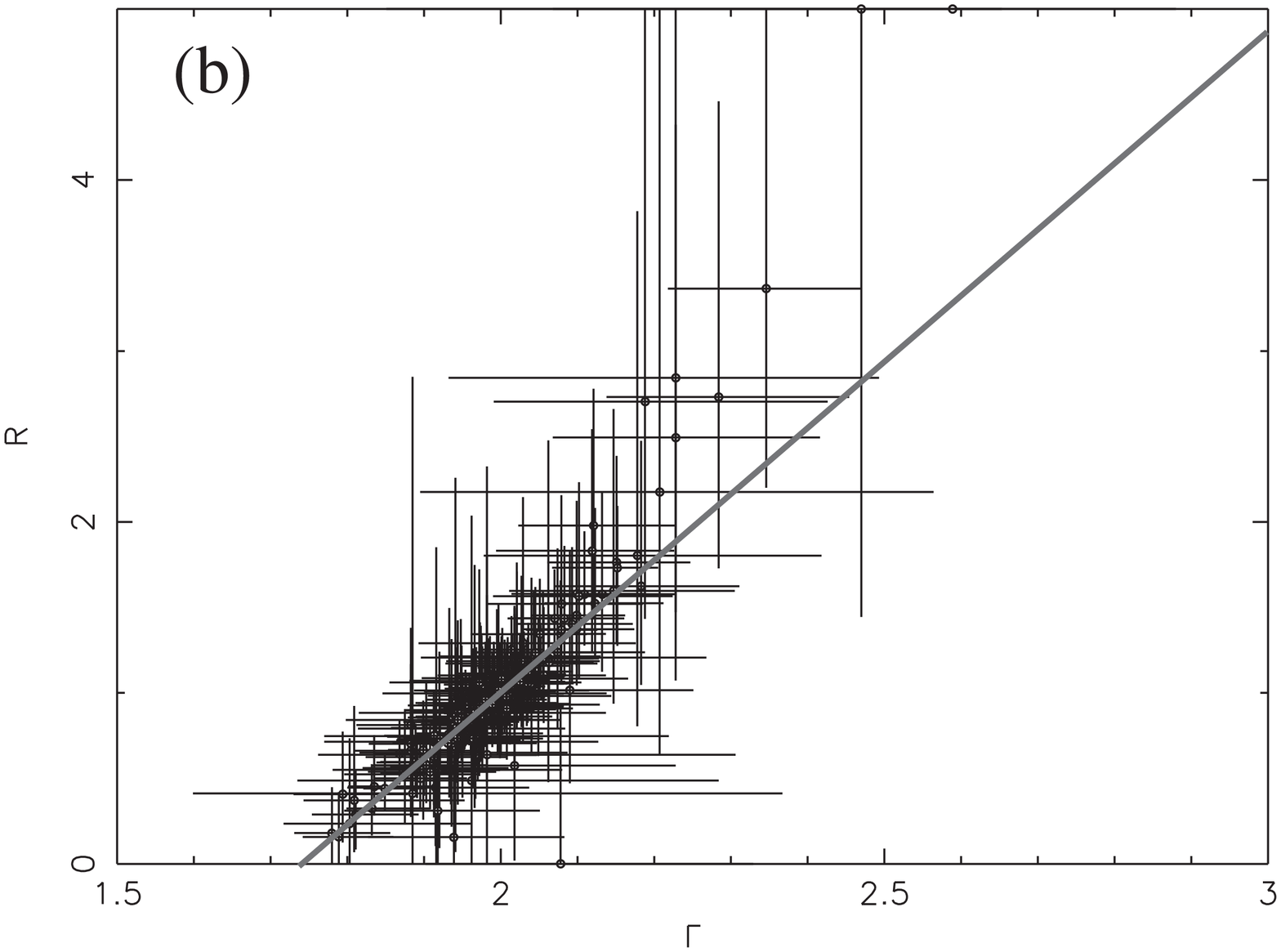}
\caption{Reflection fraction ($R$) versus powerlaw photon index ($\Gamma$) for 		
	the Seyfert 1/1.2 sample (a) and for the Monte Carlo simulations
	(b).  In both plots, the line shows the best-fit linear model for
	the Monte Carlo simulations.}
	\label{fig2}
\end{figure}

Looking at the iron line equivalent width ($EW$) and $\Gamma$, we find a
complex relationship with a ``hump'' peaking near $\Gamma \sim 2.0$
(Figure~\ref{fig1}a).  The $EW$-$\Gamma$ plot shows a correlation for
$\Gamma \lesssim 2.0$ and an anti-correlation for $\Gamma \gtrsim 2.0$,
with a peak near $\Gamma \sim 2.0$ with $EW \sim 250$ eV.  We also find
a strong correlation between $R$ and $\Gamma$ (Figure~\ref{fig2}a), with
a best-fit line of $R = -0.87 + 0.54 \: \Gamma$ ($\chi^2 = 506/349 =
1.46$).

We performed a Monte Carlo simulation to determine if our results were
an artifact of modeling degeneracies.  Each spectrum in the Monte Carlo
sample was simulated with $N_H$=$10^{22}$ cm$^{-2}$, $\Gamma$=2.0,
$R$=1.0, $E_{Fe}$=6.4 keV, and $\sigma_{Fe}$=0.23 keV. The flux and
exposure times were randomly varied for each spectrum.  The flux was
varied by randomly choosing $A$ from a uniform distribution between
0.004 and 0.06 photons keV$^-1$ cm$^-2$ s$^-1$.  The exposure time was
randomly generated from a uniform distribution between 300 and 11000s. 
The ranges for $A$ and the exposure time represent the range of $A$ and
exposure for the spectra in the full sample.

We generated 200 spectra: 100 simulated using an \RXTE\ Epoch 3
response, 50 using an Epoch 4 response, and 50 using an Epoch 5
response, roughly corresponding to our \RXTE\ sample. Each spectrum was
then fitted to the same model as our full sample.  The $R$ over $\Gamma$
plot (Figure~\ref{fig2}b) clearly shows a strong correlation with a
best-fit line of $R=-7.3+4.1 \: \Gamma$ ($\chi^2 = 28.96/159 =
0.182$), which strongly suggests that the observed $R$-$\Gamma$
correlation is a result of modeling degeneracies.  The correlation shows
a much steeper relationship than the Seyfert 1 data, due to the
large number of Seyfert 1 spectra showing $R \sim 0$.

$EW$ and $\Gamma$, however, do not suffer the same degeneracies, which
is clear from the Monte Carlo results (Figure~\ref{fig1}b).  Based
on the lack of correlation in our Monte Carlo results, we are confident
that the shape of the $EW$-$\Gamma$ plot for the data sample is real.

\begin{figure}
\plotone{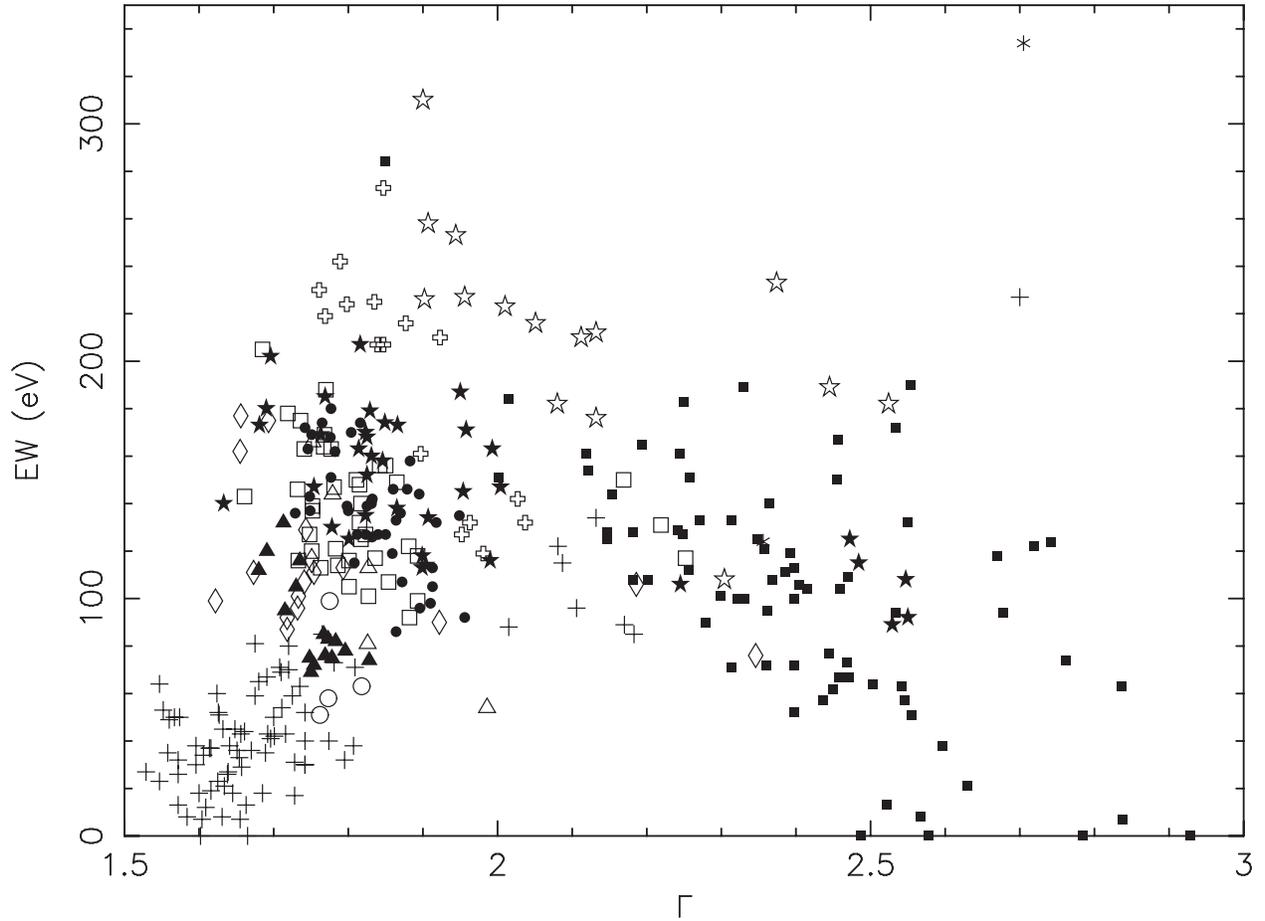}
\caption{The iron line equivalent width in eV versus the powerlaw
  	photon index.  This plot is similar to the left panel in
  	Figure~\ref{fig1}, but with each galaxy plotted with a separate
  	symbol. The open circles are 3C 111, open squares are 3C120,
  	pluses (+) are 3C273, open triangles are 3C 382, open diamonds 3C
  	390.3, open stars Akn 120, open crosses Fairall 9, filled circles
  	IC 4329A, filled squares MCG $-$6-30-15, filled triangles Mkn 509,
  	filled stars NGC 7469, and asterisks (*) TON S180.}
	\label{fig3}
\end{figure}

To further examine the $EW$-$\Gamma$ relationship, we reproduced the
$EW$-$\Gamma$ plot to show the contribution from each galaxy
(Figure~\ref{fig3}). The radio-loud galaxies form the rising leg, with
the quasar, 3C 273, anchoring the low $\Gamma$-low $EW$ portion of the
plot. The Seyfert 1 (radio quiet) and 1.2 galaxies tend to congregate at
the peak and the falling leg of the plot.  The one narrow-line Seyfert 1
diverges from the main cluster of points.

Finally, we examined $EW$ as a function of the intrinsic 2-10 keV X-ray
luminosity ($L_x$), using $H_0 = 70$ km s$^{-1}$ Mpc$^{-1}$.  We fitted
the data for each galaxy, each type, and the sample as a whole to linear
and powerlaw models.  The data were well-fit for either model. For
consistency with other publications, we report here the powerlaw
results.  For the sample as a whole, we see an anticorrelation, i.e. the
X-ray Baldwin effect \citep{iwasawa93}, with $EW \propto
L_x^{-0.14\pm0.01}$.  When examining galaxy types, however, the
anticorrelation does not always hold up (Table~\ref{table:sources}).  We
find an anticorrelation in the radio loud galaxies and the Seyfert 1.2s,
but a marginal correlation for the quasar and radio quiet Seyfert 1s.

\section{Discussion} \label{section:discussion}

\subsection{$EW$-$\Gamma$ Relationship} \label{section:deiscussion:ewg}

The simulations of \citet{george91} for the observed spectrum from an
X-ray source illuminating a half-slab showed that the spectra should
include a ``Compton hump'' and an iron line.  They found that the iron
line EW should decrease as the spectrum softens.  This is easy to
understand, since as the spectrum softens ($\Gamma$ increases), there
are fewer photons with energies above the iron photoionization
threshold. Our results show that the relationship between $EW$ and
$\Gamma$ is not quite so simple. We find a correlation between $EW$ and
$\Gamma$ when $\Gamma \lesssim 2$ and an anticorrelation when $\Gamma
\gtrsim 2$.  Other researchers have found a correlation for Seyfert 1
samples \citep{perola02, lubinski01}, but the galaxies in their samples
primarily fell in the $\Gamma \lesssim 2$ region.  \citet{page04} also
find that their data suggest a slight correlation for a sample of radio
loud and radio quiet Type 1 AGN.

A close examination of our $EW$-$\Gamma$ plot shows that the data for
different galaxy types progresses across the plot.  The plot is anchored
at the low-$\Gamma$, low-$EW$ end by the quasar, 3C 273, in our sample.
The rising arm of the plot, $\Gamma \sim 1.5 - 2.0$ and $EW \sim 0 -
300$ eV, is primarily formed by radio loud Seyfert 1 galaxies.  The
radio-quiet Seyfert 1 galaxies cluster near the $\Gamma \sim 2.0$, $EW
\sim 300$ eV peak of the hump, and the radio-quiet Seyfert 1.2 galaxies
form the falling arm of the plot for $\Gamma > 2.0$.

Physically, the most obvious difference between these sources is the
presence or absence of a strong jet.  We propose that this relationship
is driven by the degree of jet-dominance of the source.  The iron line
features are associated with the X-ray emission from the disk.  Since
the disk is essentially isotropic, it will excite an observable iron
line from matter out of our line-of-sight.  On the other hand, the jet
is beamed away from the obvious configurations of matter in the system
and, more importantly, is beamed toward us in the quasar and radio-loud
sources.  Both of these jet-related phenomena reduce the observed
equivalent width of any iron line emission associated with the jet
continuum.

In order for the $\Gamma$ to increase as the jet-dominance decreases,
the jet in these sources must have a hard X-ray component, which implies
that the radio-loud Seyferts in our sample are to be associated with
low-peaked BL Lac objects (LBLs).  BL Lac objects show two broad peaks
in their spectral energy density plots \citep{giommi94}, with the
lower-energy peak due to synchrotron emission and the higher-energy peak
due to inverse Compton emission.  BL Lacs are divided into two classes,
depending on where the peaks occur: high-peaked BL Lacs (HBLs) and LBLs.
The X-ray continuum in the HBLs is rather soft, since we are seeing the
synchrotron spectrum cutting off in these sources.  LBLs, on the other
hand, tend to have a harder X-ray continua, since we are observing well
into the inverse Compton part of the spectrum \citep{donato05}.

We also note that much of the falling arm of the $EW$-$\Gamma$
relationship is formed by MCG $-$6-30-15. Recent observations of MCG
$-$6-30-15 by Suzaku have shown that the reflection component, including
the iron line, remains relatively constant \citet{miniutti06}.  We would
expect, then, that as $\Gamma$ increases, the $EW$ should decrease,
which is exactly what we see in our data.

\subsection{$R$-$\Gamma$ Relationship} \label{section:discussion:rg}

Significant degeneracies between the photon index, absorbing column, and
reflection fraction can easily lead to false conclusions about spectral
correlations.  These degeneracies occur as these three parameters trade
off against each other in the modeling process, an effect that is
especially strong in the \RXTE\ bandpass.  Our $R$-$\Gamma$ plot shows a
strong correlation which is mimicked in our Monte Carlo results.  The
few points that lie under the main concentration are likely to be
outliers, and not indicative of a subclass of galaxy.  These points all
come from spectra that have been fitted to have $N_H = 0$, and are
primarily radio-loud galaxies.  We conclude that the observed
$R$-$\Gamma$ correlation in our sample cannot be trusted as a real
correlation.

\subsection{$EW$-$L_x$ Relationship} \label{section:discussion:ewlx}

Looking at the $EW$-$L_x$ relationship, we do see the X-ray Baldwin
effect for our sample as a whole, with a slighly shallower
anticorrelation than reported elsewhere.  We find $EW \propto
L_x^{-0.14}$, whereas \citet{iwasawa93} and \citet{jiang06} find $EW
\propto L_x^{-0.20}$ and  \citet{page04} find $EW \propto L_x^{-0.17}$.
However, when \citet{jiang06} exclude the radio loud galaxies from their
sample, they find $EW \propto L_x^{-0.10}$.  

We find, though, that when we examine our data on a galaxy-by-galaxy or
type-by-type basis, the effect is not consistent from source to source. 
At this point, we cannot determine if these variations are real or are
simply due to the small number of spectra for some of our galaxies and
types.

\section{Conclusions} \label{section:conclusions}

We have examined time-resolved spectra of 12 Seyfert 1 and 1.2 galaxies
observed by \RXTE\ over seven years. We find a complex relationship
between the iron line equivalent width and the continuum slope, with a
correlation for $\Gamma \lesssim 2$ that turns over to an
anticorrelation for $\Gamma \gtrsim 2$.  We propose that this
relationship is a possible diagnostic for jet- versus disk-dominated
sources, where jet-dominated sources show a correlation between $EW$ and
$\Gamma$, and disk-dominated sources show an anticorrelation. We also
see a strong correlation between $\Gamma$ and $R$ which is likely an
artifact of modeling degeneracies caused by the interplay of $\Gamma$,
$R$, and $n_H$ in the \RXTE\ bandpass.  Finally, we observe the X-ray
Baldwin effect for the sample as a whole, but not for each galaxy and
galaxy type individually.

\acknowledgments
This research has made use of data obtained from the High Energy
Astrophysics Science Archive Research Center (HEASARC), provided by
NASA's Goddard Space Flight Center.

This research has also made use of the NASA/IPAC Extragalactic Database
(NED) which is operated by the Jet Propulsion Laboratory, California
Institute of Technology, under contract with the National Aeronautics
and Space Administration. 

CSR gratefully acknowledges support from the National Science
Foundation under grants AST0205990 and AST0607428.

\clearpage

\begin{deluxetable}{lrrrrr}
\tablecolumns{6}
\tablecaption{Sample of \RXTE-observed Seyfert 1 and 1.2 galaxies\tablenotemark{a}}
\tablewidth{0pt}

\tablehead{
\colhead{Galaxy} &
\colhead{Seyfert} &
\colhead{Fitted} &
\colhead{Average} & 
\multicolumn{2}{c}{$EW/L_x$ correlation\tablenotemark{e}} 
\\
\cline{5-6}
\\
\colhead{} &
\colhead{Type\tablenotemark{b}} &
\colhead{Spectra\tablenotemark{c}} &
\colhead{$\sigma_{Fe K\alpha}$\tablenotemark{d}} &
\colhead{$\alpha$} &
\colhead{WV/Num.}
}

\startdata
\medskip 
All				& 		& 		& 		& -0.14$^{+0.01}_{-0.01}$	& 700/350
\\
\
\medskip 
Quasars			& 		& 		& 		& +0.09$^{+0.20}_{-0.25}$	& 105/81
\\ 
\smallskip \hspace{12pt} 
3C 273   		&  1 	&  81	& 0.329	& +0.09$^{+0.20}_{-0.25}$	& 105/81
\\

\medskip 
Broadline Seyfert 1s & 	&		&		& -0.24$^{+0.14}_{-0.15}$	& 48.0/66
\\
\smallskip \hspace{12pt} 
3C 111   		&  1 	&  4	& 0.239	& +0.70$^{+2.60}_{-1.52}$	& 0.654/4
\\
\smallskip \hspace{12pt} 
3C 120\tablenotemark{f}& 1 & 40 & 0.261	& -0.70$^{+0.63}_{-0.61}$	& 20.9/39
\\
\smallskip \hspace{12pt} 
3C 382   		&  1 	&  5	& 0.328	& -0.80$^{+1.69}_{-1.70}$	& 2.54/5
\\
\smallskip \hspace{12pt} 
3C 390.3  		&  1 	&  17	& 0.203	& -0.51$^{+0.44}_{-0.41}$	& 2.70/17
\\

\medskip 
Seyfert 1s (Radio quiet)& & 	& 		& 0.01$^{+0.30}_{0.30}$		& 23.6/31
\\
\smallskip \hspace{12pt} 
Ark 120   		&  1 	&  15	& 0.197	& -0.66$^{+0.58}_{-0.57}$	& 6.62/15
\\
\smallskip \hspace{12pt} 
Fairall 9  		&  1 	&  16	& 0.155	& +0.41$^{+0.44}_{-0.44}$	& 11.1/16
\\

\medskip 
Seyfert 1.2s	& 		& 		& 		& -0.08$^{+0.03}_{-0.03}$	& 192/169
\\
\smallskip \hspace{12pt} 
IC 4329A  		&  1.2 	&  41	& 0.214	& -0.55$^{+0.36}_{-0.37}$	& 27.5/41
\\
\smallskip \hspace{12pt} 
MCG -6-30-15 	&  1.2 	&  75	& 0.292	& -0.65$^{+0.34}_{-0.33}$	& 89.2/75
\\
\smallskip \hspace{12pt} 
Mkn 509   		&  1.2 	&  16	& 0.102	& -0.52$^{+0.91}_{-0.99}$	& 7.57/16
\\
\smallskip \hspace{12pt} 
NGC 7469  		&  1.2 	&  37	& 0.145	& -0.58$^{+0.30}_{-0.31}$	& 17.7/37
\\

\medskip 
Narrow Line Seyfert 1 &	&		& 		& 8.80$^{+20.80}_{-6.08}$	& 0.196/3
\\
\smallskip \hspace{12pt} 
TON S180 		&  1.2 	&  3	& 0.379	& 8.80$^{+20.80}_{-6.08}$	& 0.196/3

\enddata

\tablenotetext{a}{The following sources were eliminated after running the
data pipeline described in the text, due to having no spectra with at
least 125,000 net photons: Mkn 110, PG 0804+761, PG 1211+143, Mkn 79,
Mkn 335, and PG 0052+251.}
\tablenotetext{b}{Seyfert type based on the NASA Extragalactic Database}
\tablenotetext{c}{Total number of spectra extracted using our data pipeline (\S~\ref{subsection:pipeline}).}
\tablenotetext{d}{The average physical width of the Fe K$\alpha$ line
for all spectra from a source when fitted to the absorbed powerlaw model
with Compton reflection and Gaussian iron line
(\S~\ref{subsection:results}).}
\tablenotetext{e}{Results of fitting the X-ray luminosity over EW plot
to a powerlaw model; e.g. $EW \propto L_x^{\alpha}$, where $L_x$ is the
2-10 keV X-ray luminosity in ergs s$^{-1}$ and $EW$ is the iron line
equivalent width in eV.}
\tablenotetext{f}{One 3C 120 spectrum shows a flare, where $L_x$
jumps by $\sim6\times$.  The number quoted above excludes this point
from the sample. If we include the flare, we find $EW \propto
L_x^{0.07(+0.18/-0.25)}$.}

\label{table:sources}

\end{deluxetable}

\clearpage

\clearpage

\clearpage

\end{document}